\documentstyle[11pt]{article}
\begin{document}
\title{On the generality of a cosmological speculation based on Heisenberg's 
Principle}
\author{Benito Hern\'{a}ndez--Bermejo}
\date{}
\maketitle
\setlength{\baselineskip}{17pt}
{\em Departamento de F\'{\i}sica Fundamental, Universidad Nacional de 
Educaci\'{o}n a Distancia. Senda del Rey S/N, 28040 Madrid, Spain.}

{\em E--mail: \/ benito@fisfun.uned.es}

\mbox{}

\mbox{}

\begin{abstract}
In a previous work [1], it was speculated that the lack of homogeneity of 
the large-scale structure of the universe may be due to quantum fluctuations 
of space in the early universe. In [1], this was argued for a Friedmann-type 
universe for which both the curvature and the cosmological constant were 
zero. Here it is shown that the same considerations are valid for arbitrary 
values of the curvature and $\Lambda$.
\end{abstract}

\mbox{}

\mbox{}


\pagebreak

One of the most compelling and fundamental problems of modern cosmology arose 
in the last decade, when a systematic redshift survey performed by Geller and 
Huchra [2] revealed that the large-scale structure of the universe is not 
homogeneous. Observations show that this inhomogeneity has two important 
features: First, the lack of homogeneity comprises very diverse scales, 
which range from the size of galaxies to very large structures of the order 
of hundreds of Mpc: As can be shown [3], the observed structure can 
only arise if the original fluctuations in the early universe are of 
small amplitude but embrace very different scales of length. Second: 
The unexpected large size of the largest irregularities. In fact, the 
extension of these structures is an open problem, since its determination is 
limited by the extent of the survey. Estimations suggest that the 
typical size of these formations could be of the order of 100 Mpc, that is, 
$\sim5$\% of the extent of the observed universe, approximately. 

Several explanations have been proposed in order to account for such 
properties (see [4--8]). However, none of the standard 
possibilities seems to account completely for the two important features 
indicated above.

In a recent work [1], I speculated that this problem could be overcome if we 
assume that the fluctuations which originated the observed distribution of 
matter were produced by position-momentum Heisenberg's principle. The reader 
is referred to the original reference for the details, which I shall not 
recall here for the sake of brevity. The main conclusions, however, 
were that the irregularities thus produced should cover a very large 
(in fact infinite) range of scales, the fluctuations being weak at every 
scale. These two properties are in accordance with the above-mentioned 
requirements of the present-day cosmological models. 

For simplicity, it can be assumed that the early universe may be well 
described by the Friedmann equation 
\begin{equation}
  \frac{\dot{a}^2}{a^2 c^2} + \frac{K}{a^2} - \Lambda = \frac{8 \pi 
  G \rho }{3c^4} \;\: ,
   \label{fri}
\end{equation}
where $a$ is an arbitrary reference length, which changes as the universe 
evolves, $\rho $ is the mass-energy density, $K$ is the curvature of the
universe and $\Lambda$ is Einstein's cosmological constant. Equation 
(\ref{fri}) is expressed in SI units. 

In particular, when both $K$ and $\Lambda$ were taken as zero, (\ref{fri}) 
becomes
\begin{equation}
  \frac{\dot{a}^2}{a^2 c^2} = \frac{8 \pi G \rho }{3c^4} 
   \label{fri0}
\end{equation}
This is a compromise among all possible models, since we can neglect 
$\Lambda$ in a first approach ($\Lambda \ll 1$), and also because the problem 
of the total mass of the universe remains unsolved and consequently we do not 
know whether it is open or closed. In an early universe composed of energy 
and a small amount of matter, which seems to be the most likely scenario 
[9, \S 1.5], we saw that $\rho \propto a^{\epsilon - 4}$ can be taken, where 
$\epsilon > 0$, $\epsilon \ll 1$. In this case the universe expands following 
the dependence
\begin{equation}
      a(t)  \propto t^{\delta + 1/2} \;\: ,
      \label{at0}
\end{equation}
where 
\begin{equation}
\delta = \frac{\epsilon }{8-2 \epsilon }
\label{ulb}
\end{equation}
is also small and positive. It was also shown that, with full generality,  
\begin{equation}
  \Delta p \propto \dot{a} \;\: ,
  \label{dp}
\end{equation}
where $\Delta p$ is the uncertainty associated to the expansion process. From 
this, together with equations (\ref{at0}) and 
\begin{equation}
    \Delta{x} \simeq \frac{ \hbar }{ \Delta p} \;\: ,
    \label{dx}
\end{equation}
it is not difficult to obtain the important result:
\begin{equation}
      \frac{\Delta x}{a} \propto t^{-2 \delta} \; \: .
      \label{f}
\end{equation}
This ratio between the size of inhomogeneity relative to any reference length 
of the universe at a given time should be indicative of the behaviour of 
these spatial perturbations due to the uncertainty principle. 
It is apparent that (\ref{f}) diverges as $t \rightarrow 0$, 
which shows that the phenomenon embraces very diverse (in fact infinite) 
scales of lenght. Making use of (\ref{dx}) and (\ref{dp}), this divergence 
can be generally expressed as:
\begin{equation}
   \lim _{t \rightarrow 0} \; a(t) \; \dot{a} (t) = 0 \;\: .
   \label{crit}
\end{equation}
This criterion constitutes a sufficient condition for the process to be 
present, and will often be employed in future calculations. It is also clear 
that these fluctuations of space might have affected the matter distribution. 
In this sense, it was shown that the perturbations are weak at every scale. 
All these features are in full accordance [1] with the requirements of the 
cosmic structure formation theories [3] for the generation of the observed 
large-scale structure of the universe [2].

The logical step after the foregoing conlusions is the analysis of how these 
results are influenced in the case of a nonzero value of the curvature $K$ 
and the cosmological constant $\Lambda$. We shall proceed to do this in the 
present work. 

\begin{flushleft}
{\bf The case $\mid K \mid \ll 1, \: \Lambda = 0$}
\end{flushleft}

As pointed out by Hawking [10, ch. 8], the universe started its expansion at 
a rate which is exceedingly close to the critical one, namely, that for which 
the curvature is 0. It is remarkable the fact that, even today, ten billion 
years after the Big Bang, the evolution of the universe does not allow, with 
the present-day technology, an unambiguous determination of the sign of $K$. 
Consequently, it seems justified to consider in detail the situation for 
which $\mid K \mid \ll 1 $, before we deal with the general problem.

We start by considering a universe purely composed of energy ($\rho 
\propto a^{-4}$). After that, we shall study the more realistic case 
$\rho \propto a^{\epsilon - 4}$ as a perturbation. This is possible since 
$\epsilon$ is small and positive, as noted before. Thus, if we write  
$\rho = \rho _0 a^{-4}$ the Friedmann equation (\ref{fri}) is:
\begin{equation}
  \dot{a}^2 + Kc^2 = \frac{8 \pi G \rho _{0}}{3 c^2 } a^{-2} \;\: .
   \label{fri4}
\end{equation}
If $K = 0$, the solution is $a(t) = \sqrt{2 \alpha} t^{1/2}$, 
where $\alpha = (8 \pi G \rho _0 /3 c^2)^{1/2}$. Since we are assuming 
$\mid K \mid \ll 1$, we can consider $a(t,K)$ as a function of the two 
variables $t$ and $K$ and expand as a Maclaurin series in $K$:
\begin{equation}
   a(t,K) = a(t,K=0) + K \left. \frac{\partial a(t,K)}{\partial K} 
   \right|_{K=0} + o(K^2) \simeq \sqrt{2 \alpha} t^{1/2} + Kf(t) \;\:.
   \label{atk}
\end{equation}
If we now substitute this expression into the Friedmann equation (\ref{fri4}) 
and neglect all terms quadratic in $K$ we are led to a first order 
differential equation for $f$:
\begin{equation}
   t \dot{f} + \frac{1}{2} f = -\frac{c^2}{\sqrt{2\alpha }} \; t^{3/2} \;\: .
\end{equation}
This is a Cauchy equation [11]. Its general solution is: 
\begin{equation}
        f = \xi t^{-1/2} -\frac{c^2}{\sqrt{8\alpha }} t^{3/2} \,\: ,
\end{equation}
with $\xi$ a real constant of integration. From (\ref{atk}) we have:
\begin{equation}
   a(t,K) = \sqrt{2 \alpha} t^{1/2} + K \xi t^{-1/2} -\frac{K c^2}{ 
   \sqrt{8\alpha }} t^{3/2} + o(K^2) \,\: .
\end{equation}
However, the physics of the problem shows that it must be $\xi = 0$, since 
$a$ cannot diverge in the limit $t \rightarrow 0$. Then we finally obtain:
\begin{equation}
   a(t,K) = \sqrt{2 \alpha} t^{1/2}  -\frac{K c^2}{\sqrt{8\alpha }} t^{3/2} 
   + o(K^2) \,\: .
   \label{asp}
\end{equation}
We can see that the first term is the dominant one as $t \rightarrow 0$. It is 
also noticeable that if $K > 0$ (closed universe) then $a(t,K)$ increases more 
slowly than in the case $K = 0$ (and vice versa if $K < 0$), as expected. 

We can now contemplate the case $\epsilon \neq 0$ as a perturbation of 
the previous one. From equations (\ref{at0}) and (\ref{ulb}) we see that the 
first term changes its dependence from $t^{1/2}$ to $t^{\delta + 1/2}$. 
Consequently, now (\ref{asp}) must become: 
\begin{equation}
  a(t,K) = k_1 (\delta) t^{\delta + 1/2} + k_2 (\delta) t^{g(\delta) + 3/2} 
  + o(K^2) \;\: ,
  \label{kk}
\end{equation}
where $k_1 (0) = \sqrt{2 \alpha} $, $k_2 (0) = -Kc^2/\sqrt{8\alpha }$ and 
$g(0) = 0$. Since $\delta$ is small and positive we can approximate $g(
\delta) = \delta g'(0) + o(\delta ^2) \simeq \beta \delta$. Then:
\begin{equation}
  a(t,K) \simeq k_1 (\delta ) t^{\delta + 1/2} + k_2 (\delta ) t^{\beta 
  \delta + 3/2} \;\: .
  \label{apf}
\end{equation}
The perturbation is, as we see, controlled by the small parameter $\epsilon$ 
(or $\delta$ equivalently). Since  $\epsilon \ll 1$, it is to be expected 
that the term $k_1 (\delta ) t^{\delta + 1/2}$ of (\ref{apf}) is still the 
dominant one as $t \rightarrow 0$. This implies that:
\begin{equation}
   \frac{1}{2} + \delta  < \frac{3}{2} + \beta \delta \Longrightarrow 
   \beta \delta > \delta - 1  \;\: .
   \label{crit2}
\end{equation}

Now we can apply criterion (\ref{crit}) to solution (\ref{apf}):
\begin{equation}
 \lim _{t \rightarrow 0} \; a(t,K) \dot{a}(t,K) = 
 \lim _{t \rightarrow 0} \; [ k_1 k_3 t^{2 \delta} + (k_1 k_4 + k_2 k_3)t^{1+
 \delta + \beta \delta} + k_2 k_4 t^{2+2 \beta \delta}] \;\: ,
\end{equation}
where $k_3 = (\delta + 1/2)k_1$, $k_4 = (\beta \delta + 3/2)k_2$ and $k_1$, 
$k_2$ are defined as in equation (\ref{kk}). If we take into account equation 
(\ref{crit2}) we observe that: {\bf i)} $1 + \delta + \beta \delta > 2 \delta 
> 0$, and {\bf ii)} $2+2 \beta \delta > 2 \delta > 0$. Consequently:
\begin{equation}
 \lim _{t \rightarrow 0} \; a(t,K) \dot{a}(t,K) = 0.
\end{equation}
Thus the phenomenon is present for $\rho \propto a^{\epsilon - 4}$ and $\mid 
K \mid \ll 1$, independently of the sign of $K$, as we wanted to show. 

\begin{flushleft}
{\bf The general problem: Arbitrary $K$ and $\Lambda$}
\end{flushleft}

In this section we shall first demonstrate that criterion (\ref{crit}) is 
indeed satisfied for an arbitrary value of the curvature $K$. Although this 
result is far more general than that in the previous section, the proof is 
rather mathematical and the underlying physics is not as apparent as in the 
foregoing development. This is why both perspectives may be taken as 
complementary. 

We start by considering the Friedmann equation (\ref{fri}) with density 
$\rho = \rho _0 a^{\epsilon -4}$, as usual:
\begin{equation}
  \dot{a}^2 + Kc^2 = \alpha ^2 a^{\epsilon -2} \;\: ,
\end{equation}
where $\alpha = (8 \pi G \rho _0 /3 c^2)^{1/2}$ as before. Now we multiply 
both sides by $a^2$ and regroup:
\begin{equation}
  a^2 \dot{a}^2 = \alpha ^2 a^{\epsilon } -Kc^2 a^2 \;\: .
  \label{ff}
\end{equation}
We can make use of (\ref{ff}) and substitute in criterion (\ref{crit}):
\begin{equation}
  \lim _{t \rightarrow 0} \; a^2(t) \dot{a}^2(t) = 
  \lim _{t \rightarrow 0} \; [ \alpha ^2 a^{\epsilon } -Kc^2 a^2 ] \;\: .
\end{equation}
But it is obvious that:
\begin{equation}
  \lim _{t \rightarrow 0} \; a(t) = 0 \;\: .
\end{equation}
Since $\epsilon > 0$ the result is:
\begin{equation}
  \lim _{t \rightarrow 0} \; a^2(t) \dot{a}^2(t) = 0 \;\: ,
\end{equation}
and the criterion is therefore satisfied. It is interesting to observe that 
this property is, in fact, a direct consequence of the structure of the 
Friedmann equation, our only assumption concerning the exponent of the 
density. Indeed, repeating the previous calculations for a density of the 
form $\rho = \rho _0 a^{p}$, with $p$ a parameter, we conclude that criterion 
(\ref{crit}) is satisfied whenever $p > -4$. In particular, fluctuations due 
to the uncertainty principle need not be present in the unrealistic case $p = 
-4$ of a universe only composed of energy. Accordingly, it can be said that 
the radiation scenario is unstable since any perturbation due to the presence 
of matter `switches' the process on. This was exactly the result found in [1] 
for the case $K=0$, which is thus generalized for all curvatures. 

We can now be concerned with the most general case [9,12] in which 
both the curvature $K$ and $\Lambda$ can take arbitrary values: 
\begin{equation}
  \frac{\dot{a}^2}{a^2 c^2} + \frac{K}{a^2} - \Lambda = \frac{8 \pi 
  G \rho }{3c^4} 
   \label{fribis}
\end{equation}

I shall employ again the most realistic expression for the density: 
\begin{equation}   
   \rho = \rho _0 a^{\epsilon - 4} \; , \;\;\;\; \mbox{with} \; \epsilon > 0, \; 
   \epsilon \ll 1
   \label{rho}
\end{equation}

As before, I shall make use of the criterion (\ref{crit}) to give a simple 
proof of the result we are interested in. For this, only the Friedmann 
equation (\ref{fribis}) and the expression for the density (\ref{rho}) have 
to be used to write: 
\begin{equation}
  a^2 \dot{a}^2 = \alpha ^2 a^{\epsilon} - (Kc^2)a^2 + ( \Lambda c^2)a^4 \;\: ,
\end{equation}
where $\alpha$ is again a constant. It is then found that:
\begin{equation}
  \lim _{t \rightarrow 0} \; a^2 \dot{a}^2 = 
  \lim _{t \rightarrow 0} \; ( \alpha ^2 a^{\epsilon} - (Kc^2)a^2 + 
  ( \Lambda c^2)a^4 ) = 0 \;\: ,
\end{equation}
since $\epsilon > 0$ and obviously $\lim _{t \rightarrow 0} \; a(t) = 0$. 
Criterion (\ref{crit}) is therefore satisfied. This generalizes the 
result found in [1] for the case $K=0$, $\Lambda = 0$. 

\mbox{}

\begin{flushleft}
{\bf Conclusion}
\end{flushleft}

This work, which complements [1], shows how the uncertainty principle could 
provide a mechanism for the generation of fluctuations of very diverse scales 
in the early universe. These fluctuations comply to the requirements of the 
up-to-date cosmological models for the generation of the observed 
large-scale structure. Curiously, the required properties for the phenomenon 
to be present seem to be, to great extent, intrinsic to the Friedmann 
equation itself and not dependent on important features of the model such as 
the curvature or the cosmological constant. Consequently, these semiclassical 
considerations may provide an interesting starting point for a possible 
elucidation of the problem of the large-scale structure of the universe. 

\mbox{}

\begin{flushleft}
{\bf Acknowledgements}
\end{flushleft}

I would like to thank Professor M\'{a}ximo Bar\'{o}n for his careful and 
constructive revision of several versions of this work.

\pagebreak
\begin{flushleft}
{\bf References}
\end{flushleft}

\begin{description}
  
  \item[1] Hern\'{a}ndez-Bermejo, B. (1996) Heisenberg's Principle: A 
      cosmological spe\-culation. {\em Speculat. Sci. Technol.\/}, {\bf 19},
      253--7.
  
  \item[2] Geller, M. J. and Huchra, J. P. (1989) Mapping the Universe. {\em 
      Science\/}, {\bf 246}, 897--903.

  \item[3] Bertschinger, E. (1994) Cosmic structure formation. {\em Physica 
      D\/}, {\bf 77}, 354--379.
  
  \item[4] Kaiser, N. (1984) On the spatial correlations 
      of Abell clusters. {\em Astrophys. J.\/}, {\bf 284}(1), L9--L12.

  \item[5] Centrella, J. M., Gallagher, J. S., Melott, A. 
      S. and Bushouse, H. A. (1988) A case-study of large-scale structure in 
      a hot model universe. {\em Astrophys. J.\/}, {\bf 333}(1), 24--53.
  
  \item[6] Ikeuchi, S. (1981) Theory of galaxy formation triggered by 
      quasar explosions. {\em Publ. Astr. Soc. Jpn.\/}, {\bf 33}(2), 211--222.

  \item[7] Shandarin, S. F. (1994) Nonlinear dynamics of the 
      large-scale structure in the universe. {\em Physica D\/}, {\bf 77}, 
      342--353.

  \item[8] Brown, W. K. (1994) A thick, rotating universe. 
      {\em Speculat. Sci. Technol.\/}, {\bf 17}(3), 186--190.
  
  \item[9] Davies, P. (1984) {\em The Accidental Universe.\/} Cambridge: 
      Cambridge University Press.

  \item[10] Hawking, S. W. (1990) {\em A Brief History of Time: from the Big 
      Bang to Black Holes.\/} New York: Bantam Books.
  
  \item[11] Ayres, F. (1962) {\em Differential Equations.\/} New York: 
      McGraw-Hill.

  \item[12] Misner, C. W., Thorne, K. S., and Wheeler, J. A. (1973) {\em 
      Gravitation.\/} San Francisco: Freeman \& Co.

\end{description}
\end{document}